\documentclass[aps,prl,twocolumn,superscriptaddress,10pt]{revtex4}
\usepackage{amsmath,amsfonts,amssymb}
\usepackage{graphicx}
\begin{document}

\title{Theory of Half-metallic Ferrimagnetism in Double Perovskites}



\author{Onur Erten}

\author{O. Nganba Meetei}

\author{Anamitra Mukherjee}

\author{Mohit Randeria}

\author{Nandini Trivedi}
\affiliation{Department of Physics, The Ohio State University}

\author{Patrick Woodward}
\affiliation{Department of Chemistry, The Ohio State University}


\date{\today}

\begin{abstract}
We present a comprehensive theory of the temperature- and disorder-dependence 
of half-metallic ferrimagnetism in the double perovskite Sr$_2$FeMoO$_6$ (SFMO)
with $T_c$ above room temperature. 
We show that the magnetization $M(T)$ and conduction electron polarization $P(T)$ 
are both proportional to the magnetization $M_S(T)$ of localized Fe spins. 
We derive and validate an effective spin Hamiltonian, amenable to large-scale 
three-dimensional simulations. We show how $M(T)$ and $T_c$ are affected by disorder, 
ubiquitous in these materials. 
We suggest a way to enhance $T_c$ in SFMO without sacrificing polarization.
\end{abstract}


\maketitle

Double perovskites (DPs) A$_2$BB$^\prime$O$_6$ are an important family of complex oxides, 
derived from the simple ABO$_3$ perovskite structure by 
a three-dimensional (3D) checkerboard ordering of B and B$^\prime$ ions.  
One of the best studied examples is Sr$_2$FeMoO$_6$
(SFMO), a half-metallic ferrimagnet with $T_c \simeq 420$K, well above 
room temperature~\cite{serrate_2007,kobayashi_98,sarma_2000}. Clearly SFMO,
and other DPs, can have enormous technological impact if their
stoichiometry and ordering can be controlled.

From a theoretical point of view, we argue that DPs are \emph{simple} 
systems for understanding metallic ferromagnetism, despite their apparent complexity.
First, in contrast to iron, 
there is a clear separation of the localized (B) and itinerant 
degrees of freedom (coming from B$^\prime$) in the DPs. Second, in contrast to the manganites,
DPs have neither Jahn-Teller distortions nor competing superexchange, 
given the large distance between $B$-sites. Third, in contrast to dilute magnetic semiconductors,
disorder is not an essential aspect of the theoretical problem.
Important early theoretical work on half-metallic
DPs includes $T$=0 electronic structure calculations~\cite{sarma_2000},
and model Hamiltonians analyzed using various mean-field theories~\cite{millis_2001,alonso_2003,brey_2006} 
and two-dimensional (2D) simulations~\cite{sanyal_2009}. 

In this Letter, we present a comprehensive theory that gives insight
into the temperature- and disorder-dependence of the magnetic properties of metallic DPs.
We make detailed comparisons with and predictions for SFMO. Our main results are:
\\
(1) We show that both the total magnetization $M(T)$ and the conduction electron polarization $P(T)$ 
at $E_f$ are proportional to the magnetization $M_S(T)$ of localized Fe spins.  
This result is significant because, while $M(T)$ is easy to measure,
it is $P(T)$ that is of crucial importance for spintronic applications.
\\
(2) Our main theoretical advance is the derivation and validation of an effective
classical spin Hamiltonian $H_{\rm eff}$ [see eq.~(\ref{eq:H_eff})] for DPs, which
differs from both the Heisenberg and Anderson-Hasegawa models~\cite{AH}. 
We show that $H_{\rm eff}$ describes the full
$T$-dependence of the magnetization $M_S(T)$, and hence that of $M(T)$ and $P(T)$.
\\
(3) We present the results of simulations of $H_{\rm eff}$ on large 3D lattices,
including disorder effects, thus going beyond all previous theoretical calculations on SFMO.
\\
(4) We compute $M(T)$ and  $T_c$, using microscopic band-structure parameters as input,
and see how these are affected by deviations from stoichiometry and by anti-site (AS) disorder, 
ubiquitous in real materials. Ours is the first theory to show that $T_c$ is insensitive
to AS disorder, in excellent agreement with experiments, even though $M(0)$ is suppressed.
\\
(5) We conclude with a novel proposal to enhance $T_c$ of SFMO
without sacrificing polarization, using a combination
of disorder and doping.

{\bf Model Hamiltonian:} For large Hund's coupling $J_H$, the Fe$^{3+}$ 
($3d^{5}$) site has a S=5/2 ``core spin'' or local moment. 
The Mo$^{5+}$ ($4d^{1}$) contributes a t$_{2g}$ electron
which hybridizes via O with the Fe t$_{2g}$ states. 
Symmetry implies that $d_{\alpha\beta}$ electrons 
delocalize only in the $(\alpha,\beta)$-plane~\cite{harris_2004}.
Thus the motion of electrons in the 3D system decouples into 
three 2D planes.
The ``double exchange'' Hamiltonian~\cite{alonso_2003,sanyal_2009} 
describing itinerant electrons interacting
with core spins is:
\begin{eqnarray}
H&=& -t\sum_{\langle i,j\rangle,\sigma}
(\epsilon_{i\sigma}d^{\dagger}_{i\downarrow}c_{j\sigma}+h.c.)
\cr
&&
-t^{\prime}\sum_{\langle j,j^\prime \rangle,\sigma}c^{\dagger}_{j\sigma}c_{j^\prime \sigma}
+ \Delta \sum_{i}d^{\dagger}_{i\downarrow}d_{i\downarrow}
\label{eq:quantum_hamiltonian}
\end{eqnarray}
Here $d_{i\sigma} \, (c_{i\sigma})$ are fermion
operators on the Fe (Mo) sites with spin $\sigma$.
At the Fe sites $i$, we choose local axes of quantization 
along ${\bf S}_i$ and Pauli exclusion prohibits an $\uparrow$ electron.
For all the Mo sites $j$, we choose the same (global) axis of quantization.
The orientation $(\theta_i,\varphi_i)$ of the classical spins ${\bf S}_i$ then
affects the Mo-Fe hopping via
$\epsilon_{i \uparrow}= - \sin(\theta_i/2)\exp(i\varphi_i/2)$ and
$\epsilon_{i \downarrow}= \cos(\theta_i/2)\exp(i\varphi_i/2)$.

\begin{figure}[!t]
\includegraphics[width=6.7cm]{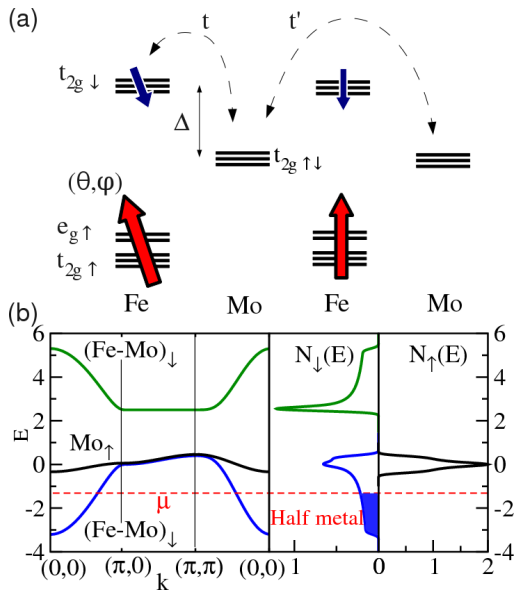}
\caption{ (a) Schematic showing energy levels at transition metal sites in
two unit cells (formula units) of SFMO. The Fe sites have localized $S=5/2$ core spins, treated as classical
vectors with orientation $(\theta,\varphi)$. The parameters $t,t^\prime$ and $\Delta$
of the Hamiltonian (\ref{eq:quantum_hamiltonian}), 
governing the dynamics of the itinerant electrons in t$_{2g}$ orbitals, are also shown. 
(b) Calculated electronic structure $E({\bf k})$ and the spin-resolved DOS $N_\uparrow(E)$ and
$N_\downarrow(E)$ in the ferromagnetic ground state with all core spins up. 
Note the half-metallic ground state in SFMO
with conduction electrons polarized opposite to core spins. 
\label{Fig:1}}
\end{figure}

The parameters in $H$ are the hopping amplitudes
$t$, between nearest neighbor (Fe-Mo) sites, and
$t^\prime$, between two Mo sites, and the charge transfer energy
$\Delta$ between Fe t$_{2g\downarrow}$ and Mo
t$_{2g}$ states; see Fig.~\ref{Fig:1}(a). The results are independent~\cite{sign_of_t}
of the sign of $t$. For now, we choose $t = 1$ as our unit of energy.
Symmetry dictates $t^\prime > 0$. 
We choose $t^\prime/t = 0.1$ and $\Delta/t = 2.5$,
using realistic band parameters for SFMO as input~\cite{sarma_2000}. 
We will show below that $t$ sets the scale for the magnetic $T_c$,
and $t = 0.27$ eV, consistent with ref.~\cite{sarma_2000},
leads to the experimental $T_c = 420$ K of SFMO.
 
\begin{figure}[!t]
\includegraphics[height=8.0cm,width=8.9cm]{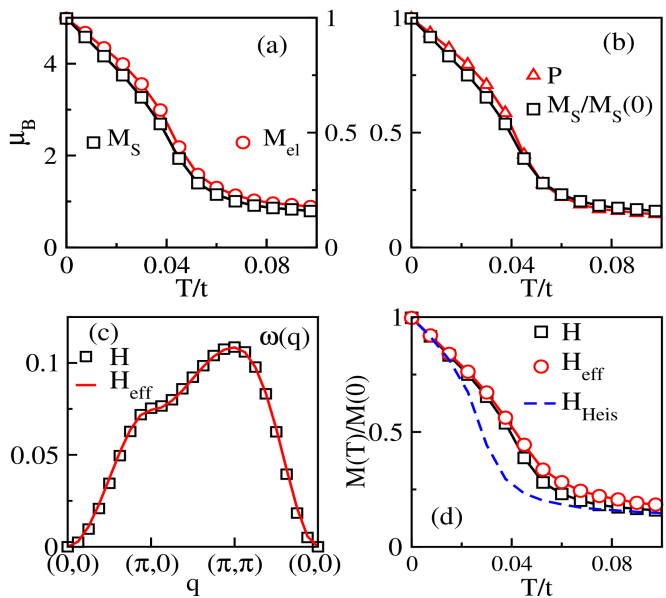}
\caption{(a) Magnetization of the core spins $M_S(T)$ and of the conduction
electrons $M_{el}(T)$ calculated using ED-MC method. (b) Conduction
electron polarization (at $E_f$) $P(T)$ and normalized $M_S(T)$. 
We see that all magnetic properties to be proportional to $M_S(T)$.
(c) Spin wave dispersion of full Hamiltonian $H$, obtained by exact diagonalization, compared
with that of the effective Hamiltonian $H_{\rm eff}$. 
Inset: Fe-Mo lattice showing nearest-neighbor ($J_1$) and next-nearest-neighbor ($J_2$)
interactions of $H_{\rm eff}$.
(d) The normalized magnetization $M(T)$ for three Hamiltonians: 
$H$, $H_{\rm eff}$ and the Heisenberg model. All results are obtained on 
$8\times 8$ systems with error bars no larger than the symbol size. 
\label{Fig:2}}
\end{figure}

We treat the quantum mechanics of ``fast'' itinerant electrons using
exact diagonalization (ED) in the background of ``slow'' S=5/2 spins, for which we
use a  $T \neq 0$ classical Monte Carlo (MC) simulation. This separation of time scales 
is justified below. Using standard ED-MC techniques, supplemented
by $T$=0 variational calculations, we obtain the results shown in 
Figs.~\ref{Fig:1} and~\ref{Fig:2}. These are limited to 2D systems for
computational reasons.

We show in Fig.~\ref{Fig:1}(b) the band structure and the spin-resolved density of states
(DOS) $N_\sigma(E)$ for the \emph{ferrimagnetic} ground state for the band-filling corresponding to SFMO
with $n=0.33$ electrons per unit cell \textit{per plane}. 
Here the conduction electrons, with magnetization $M_{el}(T=0) = 1\mu_B$ (per unit cell),
are polarized opposite to the Fe spins, with core spin magnetization $M_S(0) = 5\mu_B$.
The \emph{half-metallic} ground state, with $N_\uparrow(0) = 0$
and $N_\downarrow(0) \neq 0$, has a net magnetization $M(0) = M_S(0) - M_{el}(0) = 4\mu_B$. 

We see from Fig.~\ref{Fig:2}(a) that $M_S(T)$, $M_{el}(T)$, and hence the 
\emph{total} magnetization $M(T) = M_S(T) - M_{el}(T)$,
all have essentially the same $T$-dependence. 
Another quantity of great interest is the conduction electron
polarization $P(T)$ \emph{at} $E_f$ which determines the tunneling magnetoresistance.
This is defined by
$P(T) = [N_\downarrow(0) - N_\uparrow(0)]/[N_\downarrow(0) + N_\uparrow(0)]$,
with $E=0$ in the DOS is measured from $E_f$.
From Fig.~\ref{Fig:2}(b), we find that $P(T)$ also follows the $T$-dependent
$M_S(T)$. 

{\bf Effective Spin Hamiltonian:}
The results of Fig.~\ref{Fig:2}(a) and (b) imply that \emph{if} we had a reliable theory for the 
core spin magnetization $M_S(T)$, we 
could understand \emph{all the magnetic properties} of SFMO, including the polarization $P(T)$.
With this motivation, we derive an effective Hamiltonian $H_{\rm eff}$ for the
core spins by generalizing the two-site Anderson-Hasegawa \cite{AH} analysis 
for manganites to double perovskites. 

To derive $H_{\rm eff}$, we find the \emph{exact solution of}
(\ref{eq:quantum_hamiltonian}) \emph{for two unit cells}.
The Hilbert space, for a given t$_{2g}$ orbital, has 
three states per unit cell: Fe t$_{2g\downarrow}$ and Mo
t$_{2g,\uparrow,\downarrow}$. The smallest accessible 
filling is $n=0.5$ (per plane), i.e., one electron in two
unit cells~\cite{two-cells}.
We analytically find the lowest eigenvalue as a function
of the angle $(\theta_i - \theta_j)$ between
spins. Working in two different geometries, we find~\cite{details}
the nearest-neighbor $(J_1)$ and next-nearest-neighbor
$(J_2)$ interaction energies; see inset in Fig.~\ref{Fig:2}(c). 
Expressing these in terms of ${\bf S}_i\cdot{\bf S}_j$,
where each ${\bf S}_i$ is a unit vector,
we obtain the effective Hamiltonian
\begin{equation}\label{eq:H_eff}
 H_{\rm eff}= {-J_1}\sum_{\langle i,j
\rangle}F_{1}\left(\mathbf{S}_i\cdot\mathbf{S}_j\right) -
J_2\sum_{\langle\langle i,j
\rangle\rangle}F_{2}\left(\mathbf{S}_i\cdot\mathbf{S}_j\right)
\end{equation}
where the functions $F_{1}(x)=8\sqrt{2+\sqrt{2 + 2x}}$ and
$F_{2}(x)=(5+\sqrt{5})\sqrt{6+2\sqrt{3+2x}}$.
Our two-unit cell analysis gives explicit expressions~\cite{details} 
for $J_1$ and $J_2$, both of which are ferromagnetic
with their scale set by the kinetic energy $t$ of delocalization. 
We emphasize that the double square-root form of $H_{\rm eff}$ 
is quite different from the (single square-root) Anderson-Hasegawa model.

Next we need to understand how
we can use $H_{\rm eff}$ going beyond the simple
two-unit cell derivation. Specifically:
(i) How can we relate  $J_1$, $J_2$
to $t,t^\prime,\Delta$ and the filling $n$?
(ii) To what extent does $H_{\rm eff}$ capture the essential physics
of the full Hamiltonian $H$?

The dependence of $J_1$ and $J_2$ on microscopic parameters
can be obtained by matching the spin-wave (SW) spectra of $H_{\rm eff}$ and $H$. 
This comparison is shown in Fig.~\ref{Fig:2}(c) along certain symmetry directions. 
We find SW dispersion for $H$ using ED to compute the energy of
electrons moving in a ``frozen'' spin-wave background. The SW analysis for $H_{\rm eff}$ is
straightforward, since for small deviations from the FM ground state, $H_{\rm eff}$
reduces to a nearest and next nearest neighbor FM Heisenberg model $H_{\rm Heis}$. 
The low energy scale of $0.1t$ for spin dynamics (see Fig.~\ref{Fig:2}(c))
justifies \emph{a posteriori} our assumption of ``slow'' spins and ``fast'' electrons, 
whose bandwidth is of order $t$ (see Fig.~\ref{Fig:1}(b)). 
The same separation of energy scales also justifies the use of $T$-\emph{independent}
exchange couplings $J_1$ and $J_2$ for all $T < T_c$, as we discuss next. 

To validate the effective classical model $H_{\rm eff}$ we show in Fig.~\ref{Fig:2}(d)
that it reproduces the magnetization $M(T)$ of the full Hamiltonian $H$
over the entire range of temperatures. We note that the corresponding
Heisenberg model $H_{\rm Heis}$ gives quite \emph{different} results,
except in the $T\rightarrow 0$ limit with small spin deviations.
In other words, the $T$-dependent magnetization
of DP's cannot be modeled by a Heisenberg Hamiltonian. But
the (highly non-linear) $H_{\rm eff}$ that we have derived gives an
excellent description of the ED-MC result for the full $H$. 

\begin{figure}[!t]
\includegraphics[width=8.6cm]{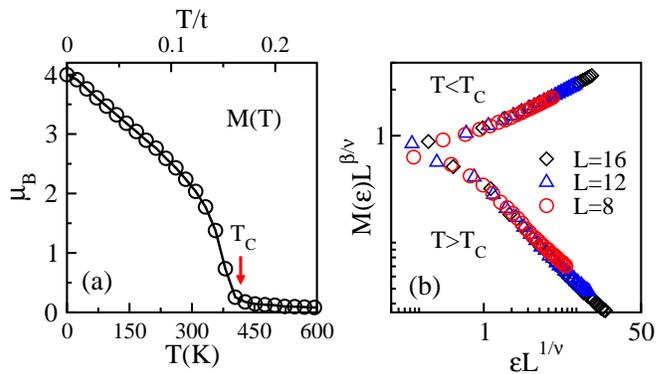}
\caption{ 
(a) Magnetization $M(T)$ from 3D simulations of $H_{\rm eff}$
on $16^3$ systems. The infinite system $T_c$ is
obtained from the scaling analysis of panel (b).
(b) Finite size scaling of data for $L^3$ systems. Here
$\varepsilon = |T - T_c|/T_c$ and $\beta=0.36$ and $\nu=0.70$ are 
the well-known 3D $O(3)$ critical exponents. 
\label{Fig:3}}
\end{figure}

The classical $H_{\rm eff}$ can be easily simulated on large 3D lattices, 
unlike the full $H$, and the results are shown in Fig.~\ref{Fig:3}. We note
the linear drop in $M(T)$ at low $T$, due to classical spin-waves,
followed by a rapid suppression of $M$ at the phase transition~\cite{2Dvs3D}.
We estimate $T_c$ in the infinite volume limit using
finite size scaling; see Fig.~\ref{Fig:3}(b).
For $t^\prime/t = 0.1$ and $\Delta/t = 2.5$, we find
$T_c = 0.14 t$. Comparing this to $T_c = 420$K for pure SFMO, we
obtain $t = 0.27$ eV, consistent with ref.~\cite{sarma_2000}.

{\bf Disorder:}
$H_{\rm eff}$ permits us to model various kinds of disorder
and deviations from stoichiometry~\cite{details}. (i) Excess Fe is modeled with
an extra spin (at a Mo site) that interacts with its neighboring
spins with a large antiferromagnetic (AF)
superexchange $S(S+1)J_{\rm AF} \simeq 34$ mev~\cite{superexchange}.
(ii) Excess Mo is modeled by removing an Fe spin from the lattice.
Both (i) and (ii) also require reevaluation of $J_1$ and $J_2$ due to
change in carrier density $n$. (iii) Here we focus on \emph{anti-site} (AS) disorder, 
the most common form of disorder in DPs, with
Fe and Mo interchanged and no change in $n$. 

\begin{figure}[!t]
\vspace{0.15cm}
\includegraphics[width=8.6cm]{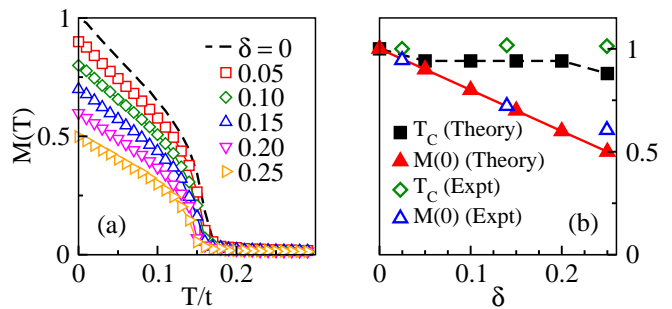}
\caption{(a) $M(T)$, \emph{normalized} by the $M(0)$ for the disorder-free system, 
for different degrees of anti-site (AS) disorder $\delta$ (see text).
(b) Comparison of theoretical results for $T_c$ and $T$=0 magnetization $M(0)$
with experiments~\cite{navarro_2003}. $M(0)$ drops like $2\delta$, while
$T_c$ is insensitive to AS disorder.}
\label{Fig:4} 
\end{figure}

We see from Fig.~\ref{Fig:4} that AS disorder systematically reduces $M(0)$
without affecting $T_c$, in excellent agreement with
experiments~\cite{ogale_1999,navarro_2003}. We quantify AS disorder using
$\delta$ the fraction of Fe atoms that are on the Mo sublattice.
The observed magnetization $M(0)[1 - 2 \delta]$ arises from
the loss of \emph{two} moments for each AS defect. One from the moment
lost at the Mo (on the Fe sublattice) and the other from
the Fe (on the Mo sublattice) antiferromagnetically coupled
to its neighbors via $J_{\rm AF}$.

There are two opposite effects of AS disorder on $T_c$ that appear to
balance each other. The strong Fe-Fe superexchange $J_{\rm AF}$ pins
the spins surrounding the Fe-defect, and makes the magnetic
order more robust against thermal fluctuations. On the other hand,
the Mo-defect leads to broken $J_1,J_2$ bonds which weaken
the magnetism. The net effect is a $T_c$ insensitive to $\delta$
for moderate levels of AS disorder, which is exactly what experiments
observe. We note that while the loss of $M(0)$ with AS disorder has
been explained earlier~\cite{ogale_1999,aguilar_2009,mishra_2011},
ours is the first theory to correctly account for $T_c$; previous
theories either found a drop~\cite{ogale_1999} in $T_c$ or an 
increase~\cite{alonso_2003}.

\begin{figure}[t]
\includegraphics[width=8.6cm]{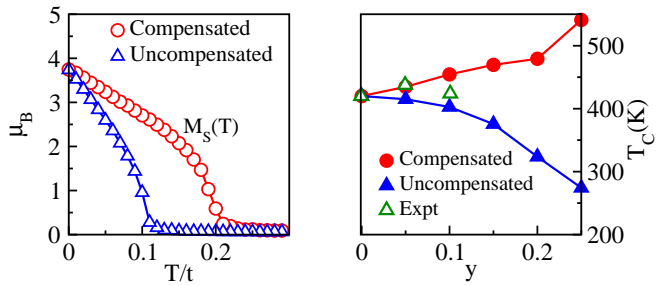}
\caption{(a) Core spin magnetization $M_S(T)$ for
La$_x$Sr$_{2-x}$Fe$_{1+y}$Mo$_{1-y}$O$_6$ for
uncompensated $(x,y)=(0,0.25)$ and compensated
$(x,y)=(0.75, 0.25)$ Fe-rich systems. 
(b) The uncompensated
$T_c(y)$ is compared with experiments \cite{topwal_2006}.
Our prediction for compensated $(x=3y)$ system shows large enhancement in $T_C$. 
}
\label{Fig:5}
\end{figure}

{\bf Raising $T_c$:}
We conclude with a proposal to raise the ferromagnetic $T_c$ without
sacrificing the conduction electron polarization $P$; see Fig.~\ref{Fig:5}. 
In brief, it involves adding excess Fe and compensating for the loss of mobile carriers
by La substitution on the Sr site.

To see how this works, let us first consider excess Fe without any
compensation: Sr$_2$Fe$_{1+y}$Mo$_{1-y}$O$_6$.
In this case, $M_S(0)$ decreases like $y$ due to the AF
alignment of excess spins. $T_c$ also decreases with $y$
because the loss of carriers $n=(1-3y)/3$ 
dominates over the enhanced pinning of moments by 
$J_{\rm AF}$ in the vicinity of defects . For the uncompensated case, the
calculated $T_c(y)$ in Fig.~\ref{Fig:5}(b) is in reasonable agreement with
experiments~\cite{topwal_2006}. 

We can compensate for the carriers by A-site substitution:
La$_x$Sr$_{2-x}$Fe$_{1+y}$Mo$_{1-y}$O$_6$
with electron density $n = (1+x-3y)/3$ per plane.
Choosing $x=3y$ counters the doping-dependent drop in $T_c$ that dominated above and we find
that $T_c$ can be significantly enhanced over that of pure SFMO due to just the 
local pinning of moments at excess Fe sites. This increase in $T_c$ goes hand-in-hand
with an unchanged $M_S(0)$; see Fig.~\ref{Fig:5}(a). 

We also note that the
polarization $P(0)$ remains 100\%.
When the Fe spins cant at finite $T$, electrons depolarize
by mixing of up and down states, however, 
no such processes are permitted at T=0, even in
a disordered Fe-rich system. We have checked using the
full Hamiltonian (\ref{eq:quantum_hamiltonian}) that,
even though the strict proportionality between $M_S(T)$ and
$P(T)$ is not observed in the presence of disorder,
the high temperature $P$ is still enhanced over the clean system
due to the large increase in $T_c$.

An alternative way to enhance $T_c$ is to add mobile electrons using
La doping, which, however, has been shown to lead to considerable
increase in AS disorder~\cite{navarro_2001}. 
In principle, compensated doping as proposed here should
introduce less AS disorder~\cite{footnote}. 

In conclusion, while we have focused here on SFMO, our theory provides a general
framework for understanding half metallic ferrimagnetism in DPs.
Interesting directions for future work include A-site substitution, Coulomb correlations on 
B$^\prime$, which may become increasingly important for larger carrier concentrations,
and spin-orbit coupling on B$^\prime$ for 5d elements.

{\bf Acknowledgments:}
Our research was supported by the Center for Emergent Materials, an NSF MRSEC (Award Number DMR-0820414).
We thank  R. Mishra, O. Restrepo, and W. Windl for discussions and 
acknowledge use of the computational facilities of the Ohio Supercomputer Center.


\begin{thebibliography}{99}
\bibitem{two-cells}
For the two unit cell geometry, $n=0.5$ is the closest we can get to the SFMO filling $n= 0.33$.
We also ignore $t^\prime$ in this geometry.
\bibitem{details}
A detailed derivation will be published elsewhere.
\end{thebibliography}
\end{document}